# A Locating Model for Pulmonary Tuberculosis Diagnosis in Radiographs

Jiwei Liu, Junyu Liu, Yang Liu, Rui Yang*, Dongjun Lv, Zhengting Cai and Jingjing Cui

*Abstract—Objective:* We propose an end-to-end CNN-based locating model for pulmonary tuberculosis (TB) diagnosis in radiographs. This model makes full use of chest radiograph (X-ray) for its improved accessibility, reduced cost and high accuracy for TB disease. *Methods:* Several specialized improvements are proposed for detection task in medical field. A false positive (FP) restrictor head is introduced for FP reduction. Anchor-oriented network heads is proposed in the position regression section. An optimization of loss function is designed for hard example mining. *Results:* The experimental results show that when the threshold of intersection over union (IoU) is set to 0.3, the average precision (AP) of two test data sets provided by different hospitals reaches 0.9023 and 0.9332. Ablation experiments shows that hard example mining and change of regressor heads contribute most in this work, but FP restriction is necessary in a CAD diagnose system. *Conclusion:* The results prove the high precision and good generalization ability of our proposed model comparing to previous works. *Significance:* We first make full use of the feature extraction ability of CNNs in TB diagnostic field and make exploration in localization of TB, when the previous works focus on the weaker task of healthy-sick subject classification.

## I. Introduction

PULMONARY tuberculosis (TB) is the ninth leading cause of death in the world, ranking first among infectious diseases and surpassing HIV/AIDS [1]. Studies show that 30 high TB burden countries, sharing nearly 85% percent of global incidence, are low-income and middle-income countries, yet progress in closing detection gap is low and big gap remains, due to lack of accurate and rapid diagnostic methods and those countries still heavily relying on sputum smear microscopy [1-2]. Because of its intrinsic limitation, the worldwide detection rate for new sputum smear-positive cases of TB stayed around 60%-70% in recent years, often unsatisfactory and insufficient for the detection of TB [2]. While chest radiograph (X-ray) could be widespread for its improved accessibility, reduced cost and high accuracy (92%) evaluated by the fourth national survey of the prevalence of TB disease [1], particularly recommended for the enhancing diagnosis of sputum smear-negative TB [3] and also scheduled to enroll the TB diagnostics pipeline by WHO.

The principal question for chest radiograph is the inter-observer and intra-observer diagnostic errors for the same image could reach up to 20-30%, with the various clinical characterization of TB, manifesting as parenchymal disease, lymphadenopathy, pleural effusion, miliary disease, or any combination thereof, and also presenting as lobar or segmental atelectasis [4]. In order to make TB diagnosis more accurate and faster, excellent computer-aided diagnosis (CAD) systems are in urgent need.

CAD system was first officially applied to medical image analysis in 2012. BV Ginneken et al. [5] proposed a fully automatic method to detect abnormalities in frontal chest radiograph which was aggregated into an overall abnormality score. In [6], the authors proposed a tuberculosis diagnosing technique combining a pixel-level textural abnormality analysis with other techniques. These initial explorations of CAD diagnosis in chest radiograph are essential, but can't meet the requirement of utility.

Recently, convolutional neural networks (CNN) have achieved remarkable results in the field of image processing [7]-[10]. The authors of [11] explored the ability of a CNN to identify different types of pathologies in chest X-ray images firstly. The obtained results demonstrated the potential of pre-trained convolutional networks as medical image feature extractors. In 2015, Bar Y et al. [12] adopted a pre-trained CNN, and a combination of traditional feature subsets PiCoDes, a compact high-level representation of low-level features (SIFTs [13], GIST, PHOG, and SSIM) as feature extractors in TB diagnosis. Mohammad et al. [14] proposed several interesting views in their paper: shallow features and ensemble models improve classification results significantly; heat maps obtained from occlusion sensitivity can be a measure of localization. Based on this, they used the features extracted from Inception [15], ResNet [16], and VGG [17] and heat maps to detect tuberculosis and achieved a 17% improvement in accuracy. Chauhan et al. [18] proposed a framework which trained a preprocessing module based on wavelet results [19] and used unsupervised pre-training to set initial weights.

Inspired by all the mentioned works, we push the target of research onto a higher level, proposing a deep neural network structure for the TB localization task instead of merely classifying the chest radiographs.

## II. Materials

This is a retrospective study that involves the datasets containing frontal chest radiographs (DICOM format) from two sources. The first dataset is provided by *Huiying Medical Technology (Beijing) Co., Ltd.* (hereinafter, *Huiying*) with labels marked by cross validation of expert doctors, and the second one is selected and provided by *Henan Provincial Chest Hospital* (hereinafter, *Henan*), where the informed consent is waived and ethical approval are obtained from the institutional review board.

Exclusion criteria are applied to remove the extreme cases: (1) less than 10 years old; (2) severe body movement leading to image distortion; (3) abnormal window width (WW) and window level (WL). A total of 5928 cases (2639 TB cases and 3289 healthy cases) from *Huiying* dataset were finally included in the study, randomly divided into three sets: training set with 1947 TB cases and 2394 healthy cases, validation set with 333 TB cases and 695 healthy cases, as well as test set with 359 TB cases and 200 healthy cases; while 279 TB cases were obtain from *Henan* dataset and served as another test set for the generalization ability of proposed model. The information and usage of the two datasets were shown in Table 1. Radiographs of TB cases in the study contained almost all the classical imaging manifestations the disease had, including airspace consolidation, miliary patterns, tuberculoma, lymphadenopathy, cavitary lesion and pleural effusion.

To obtain the regions of interest (ROI) we uploaded all chest radiographs data to Radcloud platform (http://radcloud.cn/; *Huiying Medical Technology Co., Ltd, Beijing, China*), then framed by one or more bounding boxes to include tuberculosis lesions but exclude normal tissues as much as possible. It should be noted that both lung fields were framed for tuberculosis lesions presenting as diffuse disease of both lungs. Such works were done by 3 experienced radiology physicians with 3 years of work experience, and reviewed by a veteran radiologist with nearly 30 years of work experience.

TABLE I
THE INFORMATION AND USAGE OF THE TWO DATASETS.

| Dateset | Age (vrs.mean±std) | Gender (M/F) | Size of the whole set |
|---|---|---|---|
| *Huiying* dataset | 48.0 ± 13.8 | 4879/1049 | 2639/3289 |
| *Henan* dataset | 51.±10.4 | 265/14 | 279/0 |

| Dateset | Size of the training set | Size of the validation set | Size of the testing set |
|---|---|---|---|
| *Huiying* dataset | 1947/2394 | 333/695 | 359/200 |
| *Henan* dataset | 0/0 | 0/0 | 279/0 |

Note: *vrs* = years; *std* = standard deviation; in the *Gender*, the value before the slash stands for the number of male, and the latter one represent the number of female; in the *size* of each set, the former value stands for the number of TB cases, and the latter one represents the number of healthy cases.

## III. Proposed Methodology

### A. Overview

The idea of proposed model comes from RetinaNet [20], and several specialized optimizations are made for the TB localization task. We adopt this light weight, unified framework, which composed of a backbone network and two attached task-specific subnets. The backbone is an off-the-self convolutional network using the architecture of feature pyramid network (FPN) [21], and responsible for generation of five convolutional feature maps mixing up features of different levels. This is essential for our situation since the pattern in the images of TB ranges from small points (calcification) to large area of shadow within the whole lung. The first subnet performs as convolutional object classifier head on the backbone's output; the second subnet called regressor head is a bounding box generator, doing the object localization task.

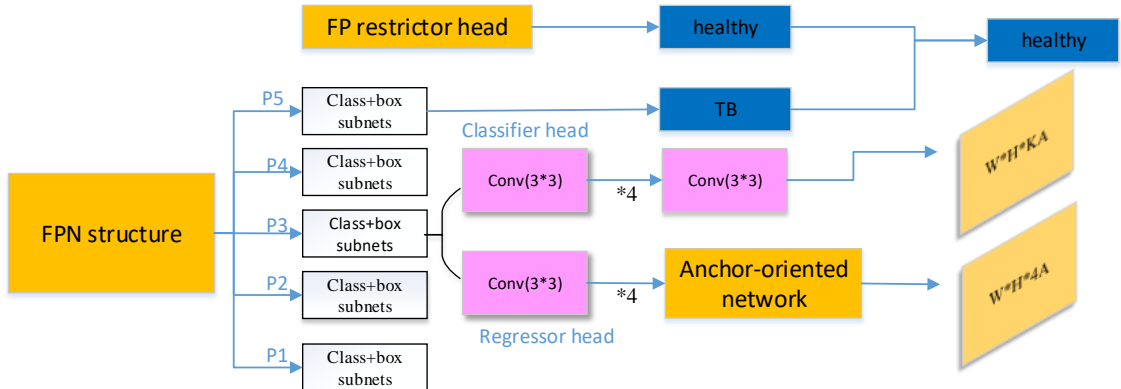

Fig. 1. The overview of proposed network architecture. The backbone of our proposed architecture is the structure of feature pyramid network (FPN structure) with five layers of feature maps (P1-P5). Each layer has two attached subsets (Class+box subnets), of which are classifier head and regressor head consisting of several convolutional kernels (Conv(3*3)) and output (W*H*KA or W*H*4A) [20]. The last convolutional kernel of regressor head is optimized as anchor-oriented network, and a false positive (FP) restrictor head is cascaded to the backbone of the network.

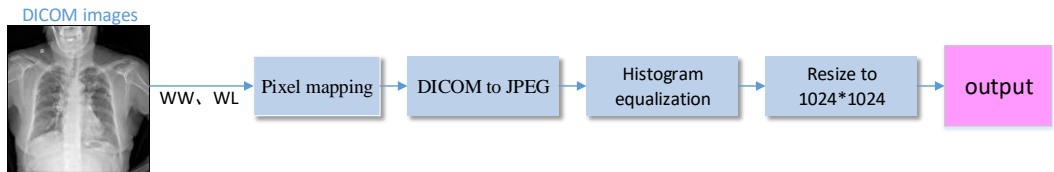

Fig. 2. The flow chart of preprocessing. The original DICOM images are transformed into the standardized JPEG format, going through the pixel mapping, format convert, histogram equalization and resizing.

In the proposed architecture, anchor-oriented network heads are adopted as the optimization of original regressor heads. And a false positive (FP) restrictor head is cascaded to the backbone of the network, considering the requirements of medical application. Meanwhile, hard sample mining (namely for samples tend to be incorrectly classified) strategy is applied to the classification loss function. The overview of proposed network is demonstrated in Fig 1.

*B. Preprocessing*

The grayscales of chest radiograph, which range widely from tens to thousands, is very likely to cause the diagnosing system to diverge. Therefore the original pixel values are mapped into the range of 0-255 according to window width (WW) and window level (WL), and then converted into Joint Photographic Experts Group (JPEG) format. In addition, original radiographs often measure as many as two thousand pixels in length, which generates a huge burden for network computation. Also considering the condition of tiny granule infections, the images are then resized to a $1024 \times 1024$ matrix. Finally, histogram equalization is performed to emphasize the features in lung without changing the grayscale in other organs and background significantly. The flow chart of preprocessing is shown in Fig 2.

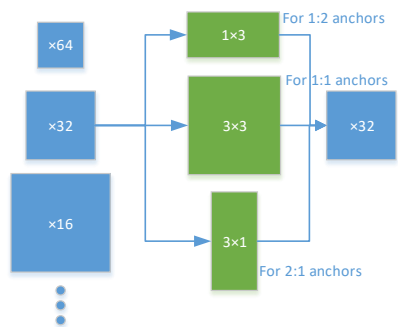

Fig. 3. Demonstration of an anchor-oriented regressor head. Blue squares in the left side are feature maps of second to last convolutional kernel in regressor head, and the blue square in the right side is output. The green blocks in the center are convolutional kernels for anchors of different aspect ratios.

*C. Anchor-oriented network heads*

Start with the success of RCNN (Regions with CNN features) [22], the structure named anchors has become the mainstream in recent works about object detection. In most studies, the convolutional kernels of output heads (regressor heads and classifier heads) were simply cascaded 3×3 squares (ignoring number of channels) and led the receptive field of all the output elements to be squares of the same size. But the anchors had different aspect ratios, and much noise could be mixed into prediction for anchors whose aspect ratios were far from 1.

As for the situation of pulmonary tuberculosis localization, the coexistence of calcification and infiltrating causes the variable sizes of ground truth to range from tens to hundreds pixels among different chest radiographs. It is also hard to predict the margins of targets basing on the inner relationship of ground truth, because lesion areas don't have a strong semantic relationship among inner pixels and the boarder. Although this kind of relationship is common in normal natural images. Therefore, much noise inside the receptive fields might affect the accuracy of pulmonary tuberculosis localization.

We half the sizes of all the anchors for the detection of calcification and other tiny targets, ensuring each ground truth matches at least an anchor in the FPN. The last convolutional kernels of regressor heads are divided into three shapes according to the aspect ratios of corresponding anchors. A 3×1 kernel is adopted for anchors whose width is shorter than height, while a 1×3 kernel is assigned to dumpy anchors. At last the predictions are concatenated by the same shape and order as original output heads. With the above adjustments, orientation differences are introduced into the output heads and the number of parameter in the heads declines slightly.

*D. FP restrictor head*

According to clinical practice, it is concluded that FN (false negative) is much more harmful in CAD diagnostic systems, so in the original detection network the parameters set is tuned to decrease FN as much as possible. As the price, more FPs are found in the results, and thus we add a high-performance FP

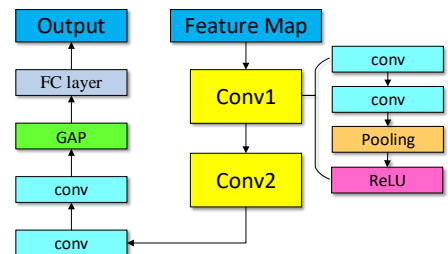

Fig. 4. FP restrictor head structure. The feature map from P5 layer of FPN is added into two classical structure VGG16(Conv1, yellow block), two convolutional layers (conv, cerulean block), a global average pooling layer (GAP, green block) and a FC layer to predict the whole image belong to a TB case or healthy case. And the VGG16 consists of two convolutional layers (conv, cerulean block), a max pooling layer (Pooling, brown block) and a ReLU activation layer (ReLU, pink block)

restrictor head for compensation. The designed head do a two-category classification of a whole radiograph to reduce the number of FP. The additional head was directly connected to P5 layer of FPN. The classical structure of VGG16 [17], which cascades two convolutional layers, a max pooling layer and a ReLU activation layer, are used twice in the new head. After that, two convolutional layers and a global average pooling (GAP) layer transfer feature map into the shape of 1×1×C, in which C is the number of channels. A fully-connected (FC) layer finally generates an output predicting whether the subject belongs to a TB case or healthy case. The loss function of this head is a simple cross-entropy loss. The logic of FP restriction is simple. If FP restrictor head outputs indicates the radiograph belongs to a healthy person, but the classifier head generates valid bounding boxes, the network will regard them as FP and ignore them.

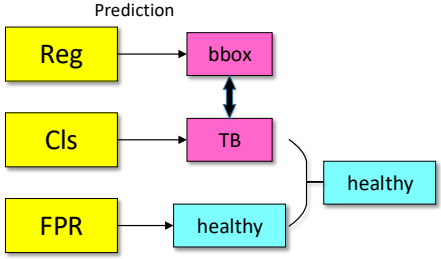

Fig. 5. Functional logic of FP restrictor head compensation. The determination of the whole radiograph (healthy or TB case) from the FP restrictor head (FPR) is to decide the retaining or removal of the valid bounding box (bbox) generated by the backbone (Cls, classifier head; Reg, regressor head).

*E. Loss function*

Tsung-Yi Lin et al. [20] proposed an original loss function named *focal loss* that significantly raises the weight of hard examples by introducing a factor $\alpha(1-p_t)^\gamma$ to the standard cross-entropy loss function. Appropriate adjustment of two parameters $\alpha, \gamma$ can reduce the weight of a large number of easy samples during training, while the scale of loss for hard samples would not be overwhelmed, which improved the accuracy of training results. The optimal combination of $\alpha, \gamma$ would be acquired after theoretical analysis and experimental demonstration.

In [20], when $\alpha$=0.25 and $\gamma$=2.0, the best detection result was achieved. After theorical analysis and experimental demonstration in the study, we found that increase of $\alpha$ causes recall rate to raise and false positive to reduce, on the contrary the decrease of $\gamma$ could reduce the false positive and raise the recall. When $\alpha$=0.40 and $\gamma$=1.0, a more balanced state could be achieved in this TB localization task.

In the TB localization task, some patterns of normal and TB radiographs are so similar that lots of hard examples emerge during training process. Reinforcement of hard example mining is done by adding an extra weight to the focal loss function. Predictive scores closer to 0.5 contribute more to the loss function. A significant increase in the recall rate of the network is observed using such loss function:

$$weight = 0.5 - (score - 0.5)^2 \quad (1)$$

$$\begin{aligned} loss_{cls} &= weight + Focal\ Loss \\ &= 0.5 - (p_t - 0.5)^2 - \alpha_t(1-p_t)^\gamma \log(p_t) \end{aligned} \quad (2)$$

Where *score* is the predictive scores in the network of one case, $loss_{cls}$ means the loss function of classifier head. The $p_t$ is defined as:

$$p_t = \begin{cases} p & , label = 1 \\ 1-p & , label = 0 \end{cases} \quad (3)$$

Because the *Focal Loss* of the classification converges quickly, and the loss of the regression converges slowly, it will cause the loss of the regression part to be larger and the loss of the classification part to be smaller when training. The final result was as follows, which could effectively reduce the probability of missed lesions:

$$loss = loss_{cls} + 0.25 \cdot loss_{reg} \quad (4)$$

Where *loss* means the total loss, the $loss_{cls}$ means the loss function of classifier head, $loss_{reg}$ means the loss function of regressor head.

As we add a FP restrictor head to the original model, the training process of each epoch is divided into two steps: first fine-tune the original detection network only using TB cases in the training set, which is the normal operation for network training. Then the FP restrictor head is trained using the whole training set when the parameters in the backbone are all frozen. The initial learning rate is set to 0.0001 and the model is trained for 50 epochs, generating a set of parameters after each epoch. The parameter set with highest average precision (AP) on validation set is adopted to the final prediction model.

## IV. STATISTICAL ANALYSIS

The work was implemented in a server with the Linux operating system (Ubuntu 16.04; Xenial, London, England) and Tensorflow v1.6 deep learning framework (https://www.tensorflow.org/), with CUDA 9.0.176 and CUDNN 7.0 (Nvidia Corporation, Santa Clara, Calif) dependencies for graphics processing unit acceleration. The computer contained an Intel Core i5-7500 3.4-gHz processor (Intel, Santa Clara, Calif), 4 TB of hard disk space, 32 GB of RAM, and a CUDA-enabled Nvidia GeForce GTX 1080 Ti graphics processing unit (Nvidia Corporation).

All of the statistical analysis and graphics were conduct using Python. The accuracy of object detection was evaluated by intersection over union (IoU), free-response receiver operating characteristic curve (fROC) and Precision-Recall curve (P-R curve) with the indicator of average precision (AP).

The definition of IoU is:

$$IoU = \frac{\text{predicted result} \cap \text{ground truth}}{\text{predicted result} \cup \text{ground truth}} \quad (5)$$

Where numerator is the area of overlap between the predicted bounding box and the ground truth bounding box, and the denominator is the area of union encompassed by both.

We define predictions with IoU larger than 0.3 as true positive (TP), since it is reasonable to mark densely aggregated targets with a large bounding box (as the condition in Fig 6). Based on the given criterion, the corresponding P-R curves and fROC curves were plotted.

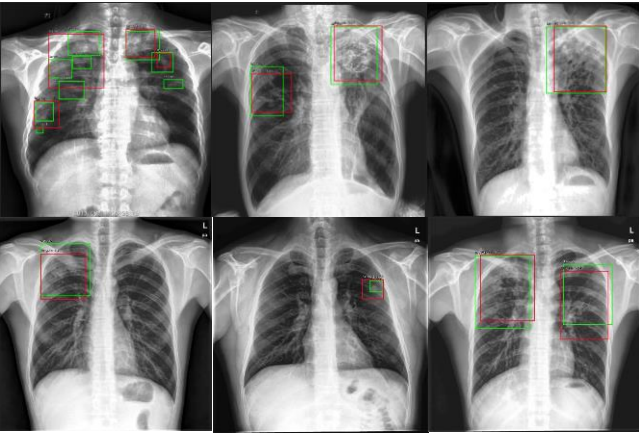

Fig. 6. Examples of Tuberculosis localization on the two test sets. The red bounding boxes are the predicted detection box and the green bounding boxes are ground truth delineated by experienced radiology physicians. The images in the first line come from *Huiying* dataset, and those in the second line are selected from *Henan* dataset.

TABLE II
THE COMPARISON OF RESULTS BETWEEN ORIGINAL RETINANET AND PROPOSED MODEL

| Model | Average precision | |
| --- | --- | --- |
|  | *Huiying* test set | *Henan* test set |
| Original RetinaNet | 0.7968 | 0.8256 |
| Proposed Method | 0.9023 | 0.9332 |

The fROC curve illustrates the relationship between sensitivity (i.e. recall) and the average number of false positives on each image at various threshold settings of the estimated classification probability, often used for evaluation of object detection tasks. The P-R curve shows the trade-off between precisions and recalls for different thresholds. AP summarizes a P-R curve as the weighted mean of precision computed at each threshold, and the weight is defined as the increase in recall from the previous threshold:

$$AP = \sum_{k=1}^{N} P_k (R_k - R_{k-1}) \qquad (6)$$

Where $N$ is the number of thresholds, $P_k$ and $R_k$ are the precision and recall at the $k$-th threshold.

V. EXPERIMENT AND RESULT

After prolonged training, the consequence of tuberculosis localization in chest radiograph are satisfactory, examples of which are shown in the Fig 6. It can be seen that the localization of TB lesions is generally accurate. For many small lesions, a large predicted bounding box tend to contain all of the lesions, and this phenomenon seemed reasonable. The final AP of our proposed model on *Huiying* test set and *Henan* test set reach 0.9023 and 0.9332 respectively, averagely increasing about 10% than the original RetinaNet. The comparison between original RetinaNet and proposed model is shown in Table 2.

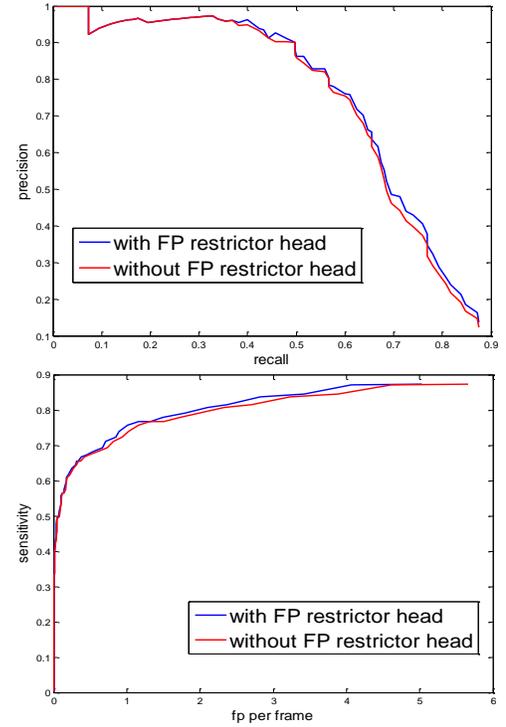

Fig. 7. The comparison of P-R curves (top) and fROC (bottom) curves between the TB localization models with and without FP restrictor head. The red curve in the figure represents the result of TB localization model without the FP restrictor head, and the blue curve represents the result of TB localization model with FP restrictor head. In the P-R curve, the horizontal is the recall rate and the vertical axis is the precision rate; in the fROC curve, the horizontal axis shows the average number of false positives on each image and the vertical axis shows the sensitivity (the same below in Figure 8 and 9).

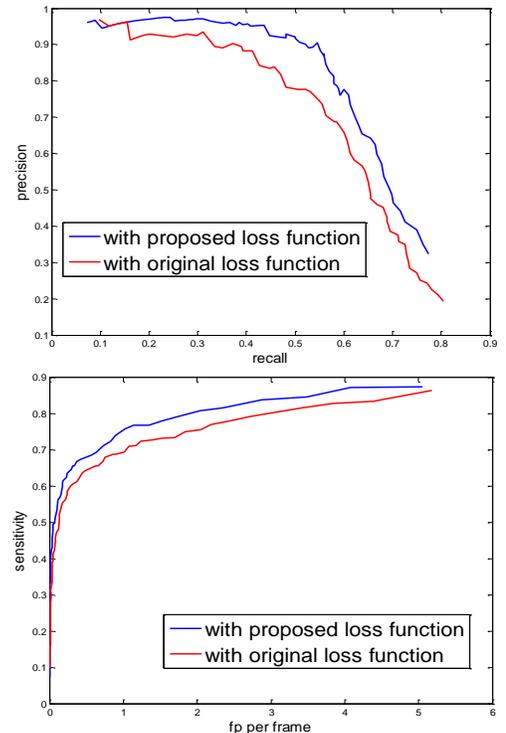

Fig. 8. The comparison of P-R curves (top) and fROC (bottom) curves between the TB localization models using newly proposed and original focal loss function. The red curve in the figure represents the result of proposed method using original loss function, and the blue curve represents the result of a model with all the condition unchanged except adopting the proposed loss function.

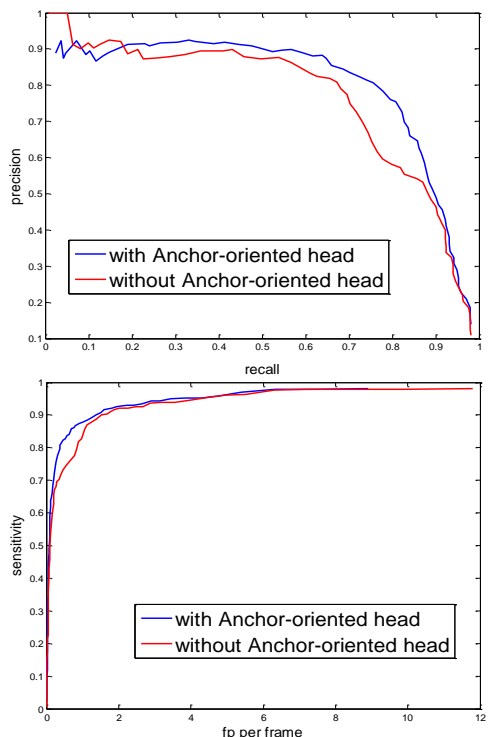

Fig. 9. The comparison of P-R curves (top) and fROC (bottom) curves between the TB localization models with and without anchor-oriented head. The red curve in the figure represents the TB localization result without the anchor-oriented heads, and the blue curve represents the TB localization result using anchor-oriented heads.

Beside the final result, we also made ablation experiments to test the contribution of each optimization to the final result.

### A. FP restrictor head

A comparative experiment to prove the validity of FP restrictor head was conducted, the performance of TB localization model with FP restrictor head slightly improved. The average precision on the *Huiying* test set increases from 0.7817 to 0.8021 and the accuracy reaches 0.95. To be intuitive, the comparison of P-R curves and fROC curves between the TB localization models with or without FP restrictor head are shown in Fig 7.

### B. Loss function

We compared the detection results of networks using proposed loss function and original loss function. AP of model with original loss function is obviously lower than that with proposed loss. When the recall is 0.6, the precision is only around 0.6 and the false positive is higher as well. The comparison of P-R curves and fROC curves between the TB localization models is shown in the Fig 8.

### C. Anchor-oriented heads

On the *Henan* test set, the effect of anchor-oriented head is examined. The comparison of the P-R curves and fROC curves is shown in Fig 9. With our proposed structure, the AP of the test set is 0.9332 and the recall of the lesion area also reaches 0.8 when FP=3. However, when abandons anchor-oriented head, AP value only reached 0.8721.

## VI. DISCUSSION

Tuberculosis occurs in the lung tissue, trachea, bronchi and tuberculous pleural, and is caused by mycobacterium tuberculosis [23]. At present, the diagnosis of tuberculosis remains a major challenge. The gold standard for clinical diagnosis of tuberculosis is the detection of mycobacterium tuberculosis in sputum or pus. Due to the slow growth of TB bacilli, it usually takes a long time, sometimes even a few months. Also, TB can be diagnosed with sputum smear microscopy, where TB bacilli in sputum samples are examined under a microscope. In addition, a skin test can be used to determine whether a patient has TB, however, a skin test is always unreliable. In general, the methods mentioned above are difficult to accept in areas where tuberculosis is endemic. Therefore, TB testing using a fast and reliable screening system, the chest X-rays, will be a critical step toward better TB diagnosis. Currently, only a few professional TB detection CAD systems are available.

In this study, the function of the automatic detection and localization to Tuberculosis through artificial intelligent in chest DR is discussed. CAD algorithms have wide applications in detecting various diseases, and they are playing an essential role as a second opinion for medical experts. Recent studies show that the deep learning method can read the appearance of various types of suspected diseases simultaneously and directly from the chest radiographs. This should be a main direction of computer-assisted detection of chest radiographs in the future.

The proposed model achieved higher average precision on two data sets collected from different sources and scanners, proving that our proposed detection network is effective and has a good generalization ability. We also tested the contribution of each optimization on the final result. Different optimization measures can improve the recall rate, accuracy rate, and decay the false positive rate of the detection network respectively.

Furthermore, our performance is reasonably close to the performance of radiologists. The accuracy on the *Huiying* test set reached 0.95, which shows that it should be possible to reach human performance in the future, or at least it can assist radiologists and public health providers in screening tuberculosis in the TB endemic areas.

Our dataset were collected respectively. Incomplete data from other clinical data of patients may wrongly improve the accuracy of the test result, but we believe that this is impossible. These TB patients are all confirmed by sputum smear or pathology in clinical. In the clinical practice, positive tuberculosis results still need to be further confirmed by clinicians, and negative results should be reviewed by radiologist to ensure the results are qualified.